%% file: main.tex
\pdfoutput=1

\documentclass[12pt,a4paper]{article}

\usepackage{ifthen}
\newboolean{pdflatex}
\setboolean{pdflatex}{true} 

\newboolean{articletitles}
\setboolean{articletitles}{true}

\newboolean{uprightparticles}
\setboolean{uprightparticles}{false}

\newboolean{inbibliography}
\setboolean{inbibliography}{false}

\def\paperauthors{LHCb collaboration} 
\def\paperasciititle{Search for excited B_c states} 
\def\papertitle{Search for excited $B_c^+$ states} 
\def\paperkeywords{{High Energy Physics}, {LHCb}} 
\def\papercopyright{CERN on behalf of the LHCb collaboration}
\def\paperlicence{CC-BY-4.0}
\def\paperlicenceurl{https://creativecommons.org/licenses/by/4.0/}

\usepackage{multirow}
\usepackage{booktabs}
\usepackage{amsmath}
\usepackage{verbatim}
\usepackage{epstopdf}
\usepackage{subfig}
\usepackage{rotating}
\usepackage{diagbox}

\input{preamble}
\input{mydefine}

\usepackage{longtable} 

\begin{document}

\renewcommand{\thefootnote}{\fnsymbol{footnote}}
\setcounter{footnote}{1}

\input{title-LHCb-PAPER}

\renewcommand{\thefootnote}{\arabic{footnote}}
\setcounter{footnote}{0}


\pagestyle{plain} 
\setcounter{page}{1}
\pagenumbering{arabic}

\input{intro}

\input{detector}

\input{selection}

\input{upperlim}

\input{acknowledgements}

\addcontentsline{toc}{section}{References}
\setboolean{inbibliography}{true}
\bibliographystyle{LHCb}
\bibliography{local,main,LHCb-PAPER,LHCb-CONF,LHCb-DP,LHCb-TDR}

\newpage
\input{LHCb_Authorship_flat_09-Oct-2017.tex}

\end{document}

%% file: preamble.tex
\usepackage[top=1in, bottom=1.25in, left=1in, right=1in]{geometry}

\usepackage{microtype}
\usepackage{lineno}  
\usepackage{xspace} 
\usepackage{caption} 

\usepackage{graphicx}  
\usepackage{color}
\usepackage{colortbl}
\graphicspath{{./figs/}} 

\usepackage{amsmath} 
\usepackage{amssymb}
\usepackage{amsfonts}
\usepackage{upgreek} 

\newcommand*\patchAmsMathEnvironmentForLineno[1]{%
\expandafter\let\csname old#1\expandafter\endcsname\csname #1\endcsname
\expandafter\let\csname oldend#1\expandafter\endcsname\csname
end#1\endcsname
 \renewenvironment{#1}%
   {\linenomath\csname old#1\endcsname}%
   {\csname oldend#1\endcsname\endlinenomath}%
}
\newcommand*\patchBothAmsMathEnvironmentsForLineno[1]{%
  \patchAmsMathEnvironmentForLineno{#1}%
  \patchAmsMathEnvironmentForLineno{#1*}%
}
\AtBeginDocument{%
\patchBothAmsMathEnvironmentsForLineno{equation}%
\patchBothAmsMathEnvironmentsForLineno{align}%
\patchBothAmsMathEnvironmentsForLineno{flalign}%
\patchBothAmsMathEnvironmentsForLineno{alignat}%
\patchBothAmsMathEnvironmentsForLineno{gather}%
\patchBothAmsMathEnvironmentsForLineno{multline}%
\patchBothAmsMathEnvironmentsForLineno{eqnarray}%
}

\usepackage{hyperxmp}

\usepackage[pdftex,
            pdfauthor={\paperauthors},
            pdftitle={\paperasciititle},
            pdfkeywords={\paperkeywords},
            pdfcopyright={Copyright (C) \papercopyright},
            pdflicenseurl={\paperlicenceurl}]{hyperref}

\usepackage[all]{hypcap} 

\input{lhcb-symbols-def} 

\usepackage{cite} 
\usepackage{mciteplus}

%% file: lhcb-symbols-def.tex
\usepackage{xspace} 
\usepackage{upgreek}

\def\lhcb {\mbox{LHCb}\xspace}
\def\atlas  {\mbox{ATLAS}\xspace}

\def\cdf    {\mbox{CDF}\xspace}

\def\tevatron {Tevatron\xspace}

\def\MagUp {\mbox{\em Mag\kern -0.05em Up}\xspace}

\ifthenelse{\boolean{uprightparticles}}%
{

 \def\Pmu         {\ensuremath{\upmu}\xspace}

 \def\Ppi         {\ensuremath{\uppi}\xspace}

 \def\Ppsi        {\ensuremath{\uppsi}\xspace}

 \def\PDelta      {\ensuremath{\Delta}\xspace}                 
 \def\PXi      {\ensuremath{\Xi}\xspace}                 
 \def\PLambda      {\ensuremath{\Lambda}\xspace}                 
 \def\PSigma      {\ensuremath{\Sigma}\xspace}                 
 \def\POmega      {\ensuremath{\Omega}\xspace}                 
 \def\PUpsilon      {\ensuremath{\Upsilon}\xspace}                 
 
 \def\PB      {\ensuremath{\mathrm{B}}\xspace}                 
                  
 \def\PD      {\ensuremath{\mathrm{D}}\xspace}

 \def\PJ      {\ensuremath{\mathrm{J}}\xspace}                 
 \def\PK      {\ensuremath{\mathrm{K}}\xspace}

 \def\Pb      {\ensuremath{\mathrm{b}}\xspace}                 
 \def\Pc      {\ensuremath{\mathrm{c}}\xspace}

 \def\Pi      {\ensuremath{\mathrm{i}}\xspace}

}
{

 \def\Pmu         {\ensuremath{\mu}\xspace}

 \def\Ppi         {\ensuremath{\pi}\xspace}

 \def\Ppsi        {\ensuremath{\psi}\xspace}                 
                  
 \mathchardef\PDelta="7101
 \mathchardef\PXi="7104
 \mathchardef\PLambda="7103
 \mathchardef\PSigma="7106
 \mathchardef\POmega="710A
 \mathchardef\PUpsilon="7107
                  
 \def\PB      {\ensuremath{B}\xspace}                 
                  
 \def\PD      {\ensuremath{D}\xspace}

 \def\PJ      {\ensuremath{J}\xspace}                 
 \def\PK      {\ensuremath{K}\xspace}

 \def\Pb      {\ensuremath{b}\xspace}                 
 \def\Pc      {\ensuremath{c}\xspace}

 \def\Pi      {\ensuremath{i}\xspace}

}

\makeatletter
\ifcase \@ptsize \relax
  \newcommand{\miniscule}{\@setfontsize\miniscule{4}{5}}
\or
  \newcommand{\miniscule}{\@setfontsize\miniscule{5}{6}}
\or
  \newcommand{\miniscule}{\@setfontsize\miniscule{5}{6}}
\fi
\makeatother

\DeclareRobustCommand{\optbar}[1]{\shortstack{{\miniscule (\rule[.5ex]{1.25em}{.18mm})}
  \\ [-.7ex] $#1$}}

\def\mup        {{\ensuremath{\Pmu^+}}\xspace}
\def\mun        {{\ensuremath{\Pmu^-}}\xspace} 

\def\cquark    {{\ensuremath{\Pc}}\xspace}

\def\bquark    {{\ensuremath{\Pb}}\xspace}

\def\pion   {{\ensuremath{\Ppi}}\xspace}

\def\pip    {{\ensuremath{\pion^+}}\xspace}
\def\pim    {{\ensuremath{\pion^-}}\xspace}

\def\kaon    {{\ensuremath{\PK}}\xspace}
  \def\Kbar    {{\kern 0.2em\overline{\kern -0.2em \PK}{}}\xspace}

\def\KorKbar    {\kern 0.18em\optbar{\kern -0.18em K}{}\xspace}

\def\Km      {{\ensuremath{\kaon^-}}\xspace}

  \def\Dbar    {{\kern 0.2em\overline{\kern -0.2em \PD}{}}\xspace}

\def\DorDbar    {\kern 0.18em\optbar{\kern -0.18em D}{}\xspace}

\def\B       {{\ensuremath{\PB}}\xspace}
\def\Bbar    {{\ensuremath{\kern 0.18em\overline{\kern -0.18em \PB}{}}}\xspace}

\def\BorBbar    {\kern 0.18em\optbar{\kern -0.18em B}{}\xspace}

\def\Bc      {{\ensuremath{\B_\cquark^+}}\xspace}
\def\Bcp     {{\ensuremath{\B_\cquark^+}}\xspace}

\def\jpsi     {{\ensuremath{{\PJ\mskip -3mu/\mskip -2mu\Ppsi\mskip 2mu}}}\xspace}

  \def\Y#1S{\ensuremath{\PUpsilon{(#1S)}}\xspace}

\def\Lbar        {{\ensuremath{\kern 0.1em\overline{\kern -0.1em\PLambda}}}\xspace}
\def\LorLbar    {\kern 0.18em\optbar{\kern -0.18em \PLambda}{}\xspace}

\def\BF         {{\ensuremath{\mathcal{B}}}\xspace}

\def\BR         {\BF}
\newcommand{\decay}[2]{\ensuremath{#1\!\to #2}\xspace}         

\def\to                 {\ensuremath{\rightarrow}\xspace}

\def\AT#1     {\ensuremath{A_{\mathrm{T}}^{#1}}\xspace}           

\def\C#1      {\ensuremath{\mathcal{C}_{#1}}\xspace}                       
\def\Cp#1     {\ensuremath{\mathcal{C}_{#1}^{'}}\xspace}                    
\def\Ceff#1   {\ensuremath{\mathcal{C}_{#1}^{\mathrm{(eff)}}}\xspace}        
\def\Cpeff#1  {\ensuremath{\mathcal{C}_{#1}^{'\mathrm{(eff)}}}\xspace}       
\def\Ope#1    {\ensuremath{\mathcal{O}_{#1}}\xspace}                       
\def\Opep#1   {\ensuremath{\mathcal{O}_{#1}^{'}}\xspace}                    

\newcommand{\tev}{\ifthenelse{\boolean{inbibliography}}{\ensuremath{~T\kern -0.05em eV}}{\ensuremath{\mathrm{\,Te\kern -0.1em V}}}\xspace}
\newcommand{\gev}{\ensuremath{\mathrm{\,Ge\kern -0.1em V}}\xspace}
\newcommand{\mev}{\ensuremath{\mathrm{\,Me\kern -0.1em V}}\xspace}
\newcommand{\kev}{\ensuremath{\mathrm{\,ke\kern -0.1em V}}\xspace}
\newcommand{\ev}{\ensuremath{\mathrm{\,e\kern -0.1em V}}\xspace}
\newcommand{\gevc}{\ensuremath{{\mathrm{\,Ge\kern -0.1em V\!/}c}}\xspace}
\newcommand{\mevc}{\ensuremath{{\mathrm{\,Me\kern -0.1em V\!/}c}}\xspace}
\newcommand{\gevcc}{\ensuremath{{\mathrm{\,Ge\kern -0.1em V\!/}c^2}}\xspace}
\newcommand{\gevgevcccc}{\ensuremath{{\mathrm{\,Ge\kern -0.1em V^2\!/}c^4}}\xspace}
\newcommand{\mevcc}{\ensuremath{{\mathrm{\,Me\kern -0.1em V\!/}c^2}}\xspace}

\def\mum  {\ensuremath{{\,\upmu\mathrm{m}}}\xspace}

\def\invfb   {\ensuremath{\mbox{\,fb}^{-1}}\xspace}

\newcommand{\stat}{\ensuremath{\mathrm{\,(stat)}}\xspace}
\newcommand{\syst}{\ensuremath{\mathrm{\,(syst)}}\xspace}

\newcommand{\chisq}{\ensuremath{\chi^2}\xspace}

\newcommand{\chisqip}{\ensuremath{\chi^2_{\text{IP}}}\xspace}

\def\gsim{{~\raise.15em\hbox{$>$}\kern-.85em
          \lower.35em\hbox{$\sim$}~}\xspace}
\def\lsim{{~\raise.15em\hbox{$<$}\kern-.85em
          \lower.35em\hbox{$\sim$}~}\xspace}

\def\sqs   {\ensuremath{\protect\sqrt{s}}\xspace}

\def\ptot       {\mbox{$p$}\xspace}
\def\pt         {\mbox{$p_{\mathrm{ T}}$}\xspace}

\def\bcvegpy    {\mbox{\textsc{Bcvegpy}}\xspace}

\def\evtgen     {\mbox{\textsc{EvtGen}}\xspace}

\def\geant      {\mbox{\textsc{Geant4}}\xspace}

\def\photos     {\mbox{\textsc{Photos}}\xspace}

\def\pythia     {\mbox{\textsc{Pythia}}\xspace}

\def\tell1  {TELL1\xspace}
\def\ukl1   {UKL1\xspace}

\newcommand{\ie}{\mbox{\itshape i.e.}\xspace}

\newcommand{\vs}{\mbox{\itshape vs.}\xspace}

%% file: mydefine.tex
\newcommand{\xx}{\ensuremath{\kern 0.5em }}

\def\BcToJpsiPi     {\decay{\Bcp}{\jpsi\pip}}

\def\Bcstar    {{\ensuremath{\B_\cquark^{*+}}}\xspace}
\def\Bctwos      {{\ensuremath{\B_\cquark(2S)^+}}\xspace}
\def\Bctwosstar  {{\ensuremath{\B_\cquark^{*}(2S)^+}}\xspace}
\def\Bctwossinglet      {{\ensuremath{\B_\cquark(2^{1}S_{0})^+}}\xspace}
\def\Bctwostriplet  {{\ensuremath{\B_\cquark(2^{3}S_{1})^+}}\xspace}
\def\Bctwosdecay  {\decay{\Bctwos}{\Bcp\pip\pim}}
\def\Bctwosstardecay  {\decay{\Bctwosstar}{\Bcstar(\to\Bcp\gamma)\pip\pim}}
\def\twoBctwos {{\ensuremath{\B_\cquark^{(*)}}(2S)^+}\xspace}
\def\twoBctwosdecay {\ensuremath{\B_\cquark^{(*)}(2S)^+ \to \B_\cquark^{(*)+} \pi^+\pi^-}}
\def\MBctwos    {\ensuremath{M(\Bctwos)}}
\def\MBctwosstar    {\ensuremath{M(\Bctwosstar)}}
\def\MBcstar    {\ensuremath{M(\Bcstar)}}
\def\MBcp    {\ensuremath{M(\Bcp)}}
\def\Mpip    {\ensuremath{M(\pip)}}
\def\Mpim    {\ensuremath{M(\pim)}}
\def\MJpsipi    {\ensuremath{M(\jpsi\pip)}}
\def\MBcpipi    {\ensuremath{M(\Bcp\pip\pim)}}

\usepackage{multirow} 
\usepackage{booktabs} 
\usepackage{rotating}

%% file: title-LHCb-PAPER.tex
\begin{titlepage}
\pagenumbering{roman}

\vspace*{-1.5cm}
\centerline{\large EUROPEAN ORGANIZATION FOR NUCLEAR RESEARCH (CERN)}
\vspace*{1.5cm}
\noindent
\begin{tabular*}{\linewidth}{lc@{\extracolsep{\fill}}r@{\extracolsep{0pt}}}
\ifthenelse{\boolean{pdflatex}}
{\vspace*{-1.5cm}\mbox{\!\!\!\includegraphics[width=.14\textwidth]{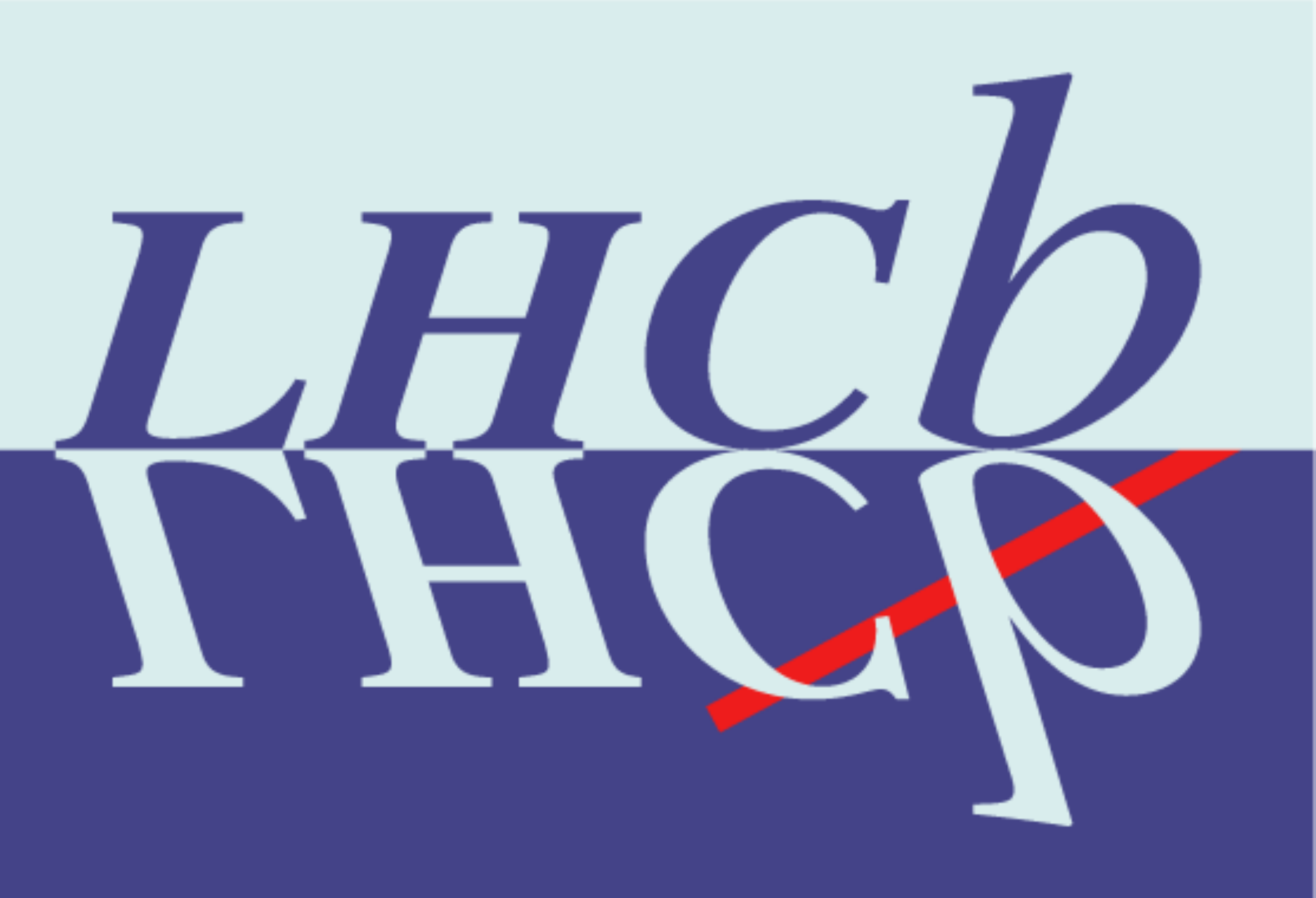}} & &}%
{\vspace*{-1.2cm}\mbox{\!\!\!\includegraphics[width=.12\textwidth]{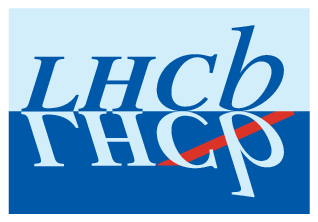}} & &}%
\\
 & & CERN-EP-2017-306 \\  
 & & LHCb-PAPER-2017-042 \\   
 & & February 12, 2018 \\ 
\end{tabular*}

\vspace*{4.0cm}

{\normalfont\bfseries\boldmath\huge
\begin{center}
  \papertitle 
\end{center}
}

\vspace*{2.0cm}

\begin{center}
\paperauthors\footnote{Authors are listed at the end of this paper.}
\end{center}

\vspace{\fill}

\begin{abstract}
  \noindent
  A search is performed in the invariant mass spectrum of the $\Bcp\pip\pim$ system
  for the excited $\Bcp$ states \Bctwossinglet and \Bctwostriplet
  using a data sample of $pp$ collisions collected by the LHCb experiment at the centre-of-mass energy of $\sqs = 8\tev$,
  corresponding to an integrated luminosity of $2\invfb$.
  No evidence is seen for either state.
  Upper limits on the ratios
  of the production cross-sections of the $\Bctwossinglet$ and $\Bctwostriplet$ states 
  times the branching fractions of $\decay{\Bctwossinglet}{\Bcp\pip\pim}$ and $\decay{\Bctwostriplet}{\Bcstar\pip\pim}$
  over the production cross-section of the $\Bc$ state
  are given as a function of their masses.
  They are found to be between 0.02 and 0.14 at $95\%$ confidence level
  for $\Bctwossinglet$ and $\Bctwostriplet$ in the mass ranges $[6830, 6890]\mevcc$ and $[6795,6890]\mevcc$, respectively.

\end{abstract}

\vspace*{2.0cm}

\begin{center}
  Published in JHEP 01 (2018) 138 
\end{center}

\vspace{\fill}

{\footnotesize 
\centerline{\copyright~\papercopyright, licence \href{\paperlicenceurl}{\paperlicence}.}}
\vspace*{2mm}

\end{titlepage}


\newpage
\setcounter{page}{2}
\mbox{~}

\cleardoublepage

%% file: intro.tex
\section{Introduction}
\label{sec:intro}
The $B_c$ meson family is unique in the Standard Model,
as its states contain two different heavy-flavour valence quarks.
It has a rich spectroscopy, predicted 
by various models~\cite{Gershtein:1987jj, Chen:1992fq, Eichten:1994gt, Kiselev:1994rc,
  Gupta:1995ps, Fulcher:1998ka, Ebert:2002pp, Godfrey:2004ya, Wei:2010zza, Rai:2006dt, El:2005,
  Berezhnoy:1997fp,Gouz:2002kk,Cheung:1995ye}
and lattice QCD~\cite{Davies:1996gi}.
The ground state of the $B_c$ meson family,
the $\Bcp$ meson, was first observed 
by the \cdf experiment~\cite{Abe:1998wi,Abe:1998fb} at the \tevatron collider in 1998.\footnote{Sums over charge-conjugated modes are implied throughout this paper.} 
Recently, the \atlas collaboration reported observation of an excited $B_c$ state with a mass of \mbox{$6842 \pm 4\stat \pm 5\syst \mevcc$}~\cite{Aad:2014laa}.
Since the production cross-section of the $\Bctwostriplet$ state is predicted to be more than 
twice that of the $\Bctwossinglet$ state~\cite{Chang:2005hq,Gouz:2002kk,Godfrey:2004ya,Gao:2010zzc},
the most probable interpretation of the single peak 
is either a signal for $\decay{\Bctwostriplet}{\Bcstar\pip\pim}$, followed by $\Bcstar \to \Bcp \gamma$ with a missing low-energy photon,
or an unresolved pair of peaks from the decays $\decay{\Bctwossinglet}{\Bc\pip\pim}$ and $\decay{\Bctwostriplet}{\Bcstar\pip\pim}$.\footnote
{The spectroscopic notation
$n^{2s+1}L_{J}$ is used, where $n$ is the radial quantum number, 
$s$ the total spin of the two valence quarks, $L$ their relative angular momentum~($S$ implies $L=0$), 
and $J$ the total angular momentum of the system, \ie spin of the excited state.
$\Bcstar$ denotes the ${\ensuremath{\B_\cquark(1^{3}S_{1})^+}}\xspace$ state.}
The ${\ensuremath{\B_\cquark(2^{1}S_{0})^+}}\xspace$ and ${\ensuremath{\B_\cquark(2^{3}S_{1})^+}}\xspace$ states
are denoted as $B_c(2S)^{+}$ and $B_c^{*}(2S)^{+}$ hereafter, 
and $\twoBctwos$ denotes either state.

In the present paper, the $\Bctwos$ and $\Bctwosstar$
mesons are searched for
using $pp$ collision data collected by the \lhcb experiment at $\sqs = 8\tev$,
corresponding to an integrated luminosity of $2 \invfb$.
The $\Bctwos$ and $\Bctwosstar$ mesons are reconstructed  
through the decays 
\mbox{$\decay{\Bctwos}{\Bcp\pip\pim}$} and \mbox{$\decay{\Bctwosstar}{\Bcstar\pip\pim}$}
with $\Bcstar \to \Bcp \gamma$, $\Bcp \to \jpsi \pip$ and $\jpsi \to \mup\mun$.
The branching fraction of the $B_c^{(*)}(2S)^{+} \to B_c^{(*)+} \pip\pim$ decay,
$\BR(B_c^{(*)}(2S)^{+} \to B_c^{(*)+} \pip\pim)$,
is predicted to be between 39\% and 59\%~\cite{Gouz:2002kk,Godfrey:2004ya}.
The low-energy photon in the $\Bctwosstar$ decay chain is not reconstructed.
The $\Bctwosstar$ state still appears in the invariant mass $\MBcpipi$ spectrum as a narrow mass peak~\cite{Gao:2010zzc,Berezhnoy:2013sla}, 
which is centered at \mbox{$\MBctwos - \Delta M$},
where 
\begin{equation}
\Delta M \equiv 
\left[ \MBcstar - \MBcp \right] -
\left[ \MBctwosstar - \MBctwos \right],
\end{equation}
and $\MBcp$ is the known mass of $\Bcp$.
According to theoretical predictions~\cite{Gershtein:1987jj, Chen:1992fq, Eichten:1994gt, Kiselev:1994rc,
  Gupta:1995ps, Fulcher:1998ka, Ebert:2002pp, Godfrey:2004ya, Wei:2010zza, Rai:2006dt, El:2005},
the mass of the $\Bctwos$ state, $\MBctwos$, is expected to be in the range $[6830,6890]\mevcc$, 
and $\Delta M$ in the range $[0,35]\mevcc$, 
such that the peak position of the $\Bctwosstar$ state in $\MBcpipi$ is expected to be in the range $[6795,6890]\mevcc$.

%% file: detector.tex
\section{Detector and simulation}
\label{sec:Detector}
The \lhcb detector~\cite{Alves:2008zz,LHCb-DP-2014-002} is a single-arm forward
spectrometer covering the \mbox{pseudorapidity} range $2<\eta <5$,
designed for the study of particles containing \bquark or \cquark
quarks. 
The detector includes a high-precision tracking system
consisting of a silicon-strip vertex detector surrounding the $pp$
interaction region, a large-area silicon-strip detector~(TT) located
upstream of a dipole magnet with a bending power of about
$4{\mathrm{\,Tm}}$, and three stations of silicon-strip detectors and straw
drift tubes placed downstream of the magnet.
The tracking system provides a measurement of momentum, \ptot, of charged particles with
a relative uncertainty that varies from 0.5\% at low momentum to 1.0\% at 200\gevc.
The minimum distance of a track to a primary vertex (PV), the impact parameter (IP), 
is measured with a resolution of $(15+29/\pt)\mum$,
where \pt is the component of the momentum transverse to the beam, in\,\gevc.
Different types of charged hadrons are distinguished using information
from two ring-imaging Cherenkov detectors. 
Photons, electrons and hadrons are identified by a calorimeter system consisting of
scintillating-pad and preshower detectors, an electromagnetic
calorimeter and a hadronic calorimeter. Muons are identified by a
system composed of alternating layers of iron and multiwire
proportional chambers.
The online event selection is performed by a trigger, 
which consists of a hardware stage, based on information from the calorimeter and muon
systems, followed by a software stage, which applies a full event
reconstruction.
At the hardware stage, events are required to have at least one muon with high $\pt$ or 
a hadron with high transverse energy. 
At the software stage, two muon tracks or three charged tracks 
are required to have high \pt and to form a secondary vertex
with a significant displacement from the interaction point.

In the simulation, $pp$ collisions are generated using
\pythia6~\cite{Sjostrand:2006za}
 with a specific \lhcb
configuration~\cite{LHCb-PROC-2010-056}. 
The generator \bcvegpy~\cite{Chang:2005hq}
is used to simulate the production of $B_c$ mesons.
Decays of hadronic particles
are described by \evtgen~\cite{Lange:2001uf}, in which final-state
radiation is generated using \photos~\cite{Golonka:2005pn}. 
The
interaction of the generated particles with the detector, and its response,
are implemented using the \geant
toolkit~\cite{Allison:2006ve} as described in
Ref.~\cite{LHCb-PROC-2011-006}.
In the default simulation,
the masses of the excited $B_c$ states are set as
$\MBctwos = 6858\mevcc$, $\MBctwosstar = 6890\mevcc$ 
and $\MBcstar = 6342\mevcc$,
corresponding to $\Delta M = 35\mevcc$,
and the $\Bctwosstar$ state is assumed to be produced unpolarised.
Simulated samples with different mass settings,
which cover the expected mass range of the $\twoBctwos$ states,
are generated to study 
variations in the reconstruction efficiency.

%% file: selection.tex
\section{Event selection}
\label{sec:selection}
\begin{figure}
  \vskip-0.5cm
 \begin{center}
    \includegraphics[scale=0.6]{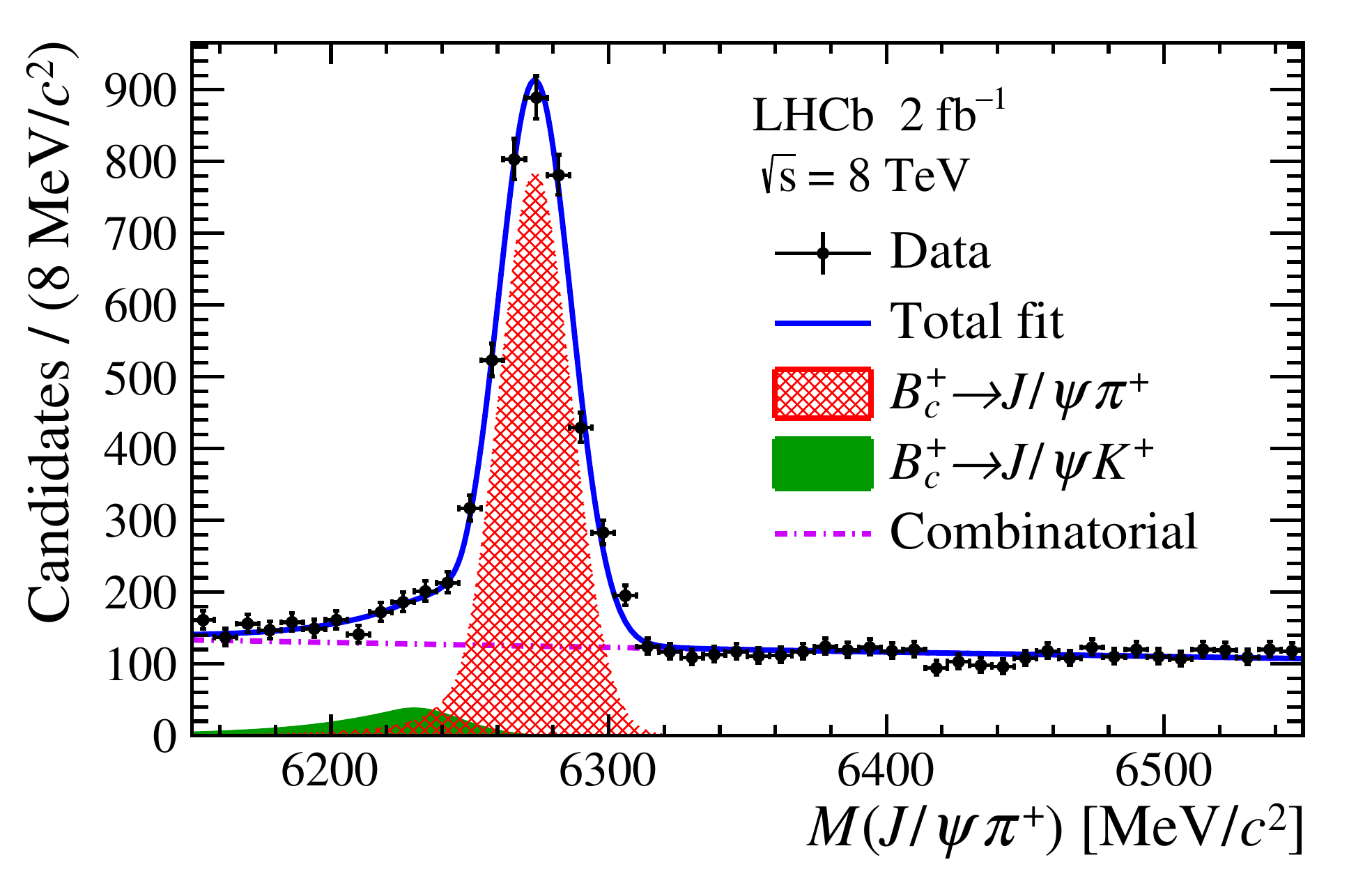}
   \caption{
   Invariant mass distribution of the selected $\Bcp \to \jpsi \pip$ candidates.
   The points with error bars represent the data.
   The blue solid line is the fit to data.
   The red cross-hatched area shows the signal.
   The green shaded area represents the $\Bc \to \jpsi K^{+}$ background.
   The violet dash-dotted line is the combinatorial background.
   }
   \label{fig:Bcyield}
 \end{center}
 \vskip-1.0cm
\end{figure}

To select $\BcToJpsiPi$ decays,
$\jpsi$ candidates are formed from pairs of opposite-charge tracks.
The tracks are required to have $\pt$ larger than
$0.55\gevc$ and good track-fit quality,
to be identified as muons,
and to originate from a common vertex.
Each $\jpsi$ candidate with an invariant mass
between $3.04\gevcc$ and $3.14\gevcc$ is combined
with a charged pion to form a $\Bcp$ candidate. 
The pion is required to have $\pt>1.0\,\gevc$
and good track-fit quality.
The \jpsi\ candidate and the charged pion are required
to originate from a common vertex,
and the $\Bcp$ candidates must have
a decay time larger than $0.2\,{\rm ps}$.
Each of the particles is associated to the PV
that has the smallest $\chisqip$,
where $\chisqip$ is defined as 
the difference in the vertex-fit $\chisq$ of a given PV reconstructed with and
without the particle under consideration.
The $\chisqip$ of the $\Bcp$~($\pip$) candidate is required to be $<25$~($>9$) with respect to the associated PV of the $\Bcp$ candidate.
To further suppress background,
a requirement on a boosted decision tree~(BDT)~\cite{Breiman,AdaBoost} classifier is applied.
The BDT classifier uses information from
the $\chisqip$ of the two muons, the pion, the $\jpsi$, and the $\Bcp$ mesons with respect to the associated PV;
the $\pt$ of both muons, the $J/\psi$ and $\pip$ mesons;
and the decay length, decay time, and the vertex-fit $\chisq$ of the
$\Bcp$ meson.
The BDT is trained with signal events taken from simulation and background events 
from the upper sideband containing $\Bcp$ candidates with masses in the range $[6370,6600]\mevcc$. 
The distributions of the BDT response for the simulation and the background subtracted data are in agreement.
The criterion on the BDT output is chosen to maximise the figure of merit $S/\sqrt{S+B}$,
where $S$ and $B$ are the expected numbers of signal and background in the range $\MJpsipi \in [6251,6301]\mevcc$.
The mass of the $\jpsi$ candidates is constrained to the known value~\cite{PDG2014}
to improve the $\Bcp$ mass resolution.\footnote{The $\jpsi$ mass is taken to be $3096.916 \mevcc$ according to the 2014 edition of the Review of Particle Physics~\cite{PDG2014}, rather than $3096.900\mevcc$ in the 2016 edition~\cite{PDG2016}.
The effect of this choice on the final result is negligible.}
The $\Bcp$ signal yield is obtained by performing an unbinned extended maximum
likelihood fit to the $\MJpsipi$ mass distribution, as shown in Fig.~\ref{fig:Bcyield}.
The signal component is modelled by a Gaussian function with asymmetric power-law tails as determined from simulation.
The mean and resolution of the Gaussian function are free parameters in the fit.
The combinatorial background is described with an exponential function.
The contamination from the Cabibbo-suppressed channel $\Bcp \to \jpsi K^+$, with the kaon misidentified as a pion, is described by a Gaussian function with asymmetric power-law tails.
The parameters are also fixed from simulation, with only the Gaussian mean related to the $\Bcp \to \jpsi \pip$ signal as a free parameter to account for the possible small mass difference in data and simulation.
The signal yield of $\Bcp$ decays is determined to be $3325 \pm 73$.

\begin{figure}
  \vskip-0.5cm
  \begin{center}
     \subfloat[MLP category: (0.02,0.2) ]{ \includegraphics[scale=0.39]{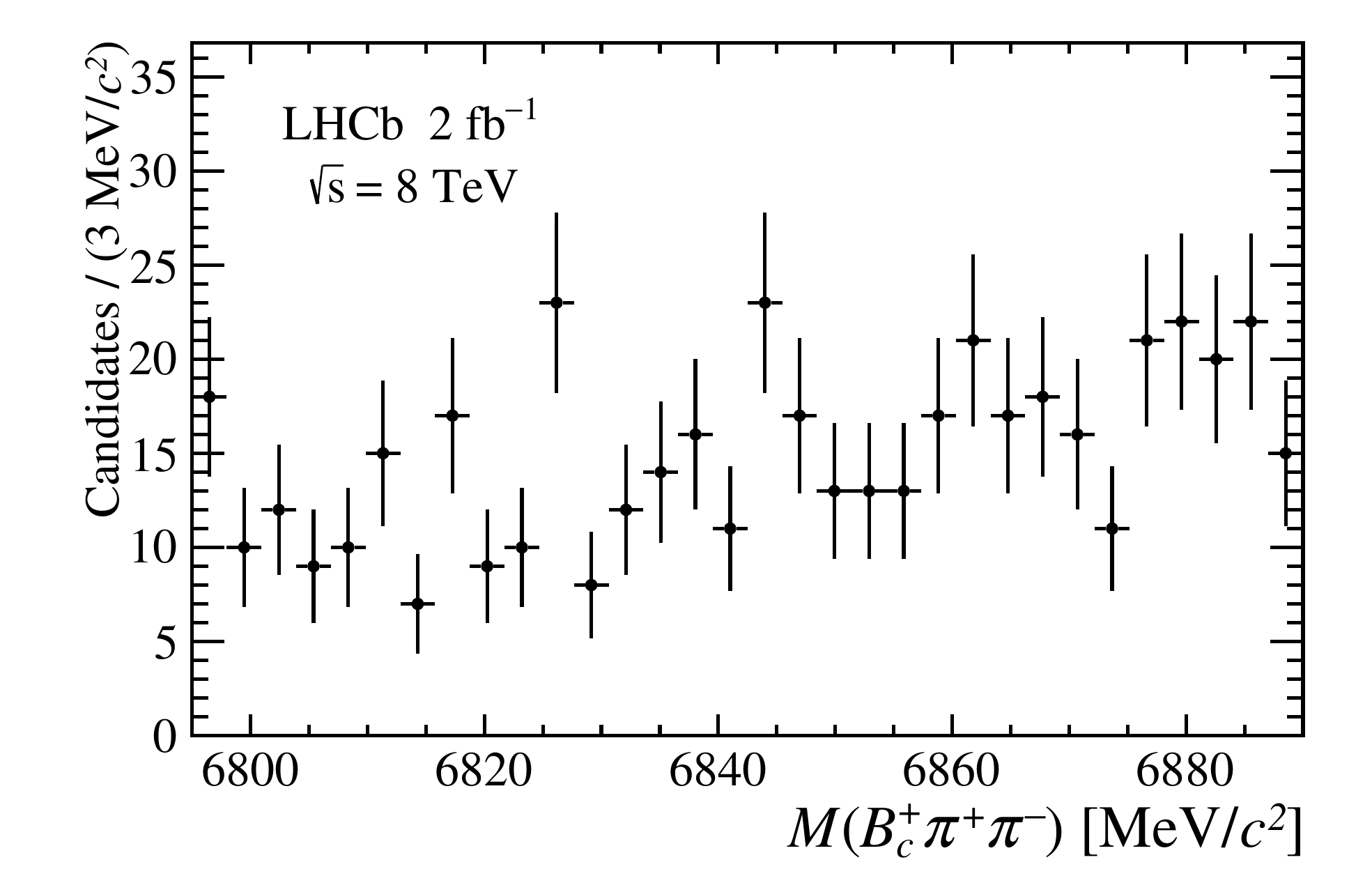} }
     \subfloat[MLP category: [0.2,0.4) ]{ \includegraphics[scale=0.39]{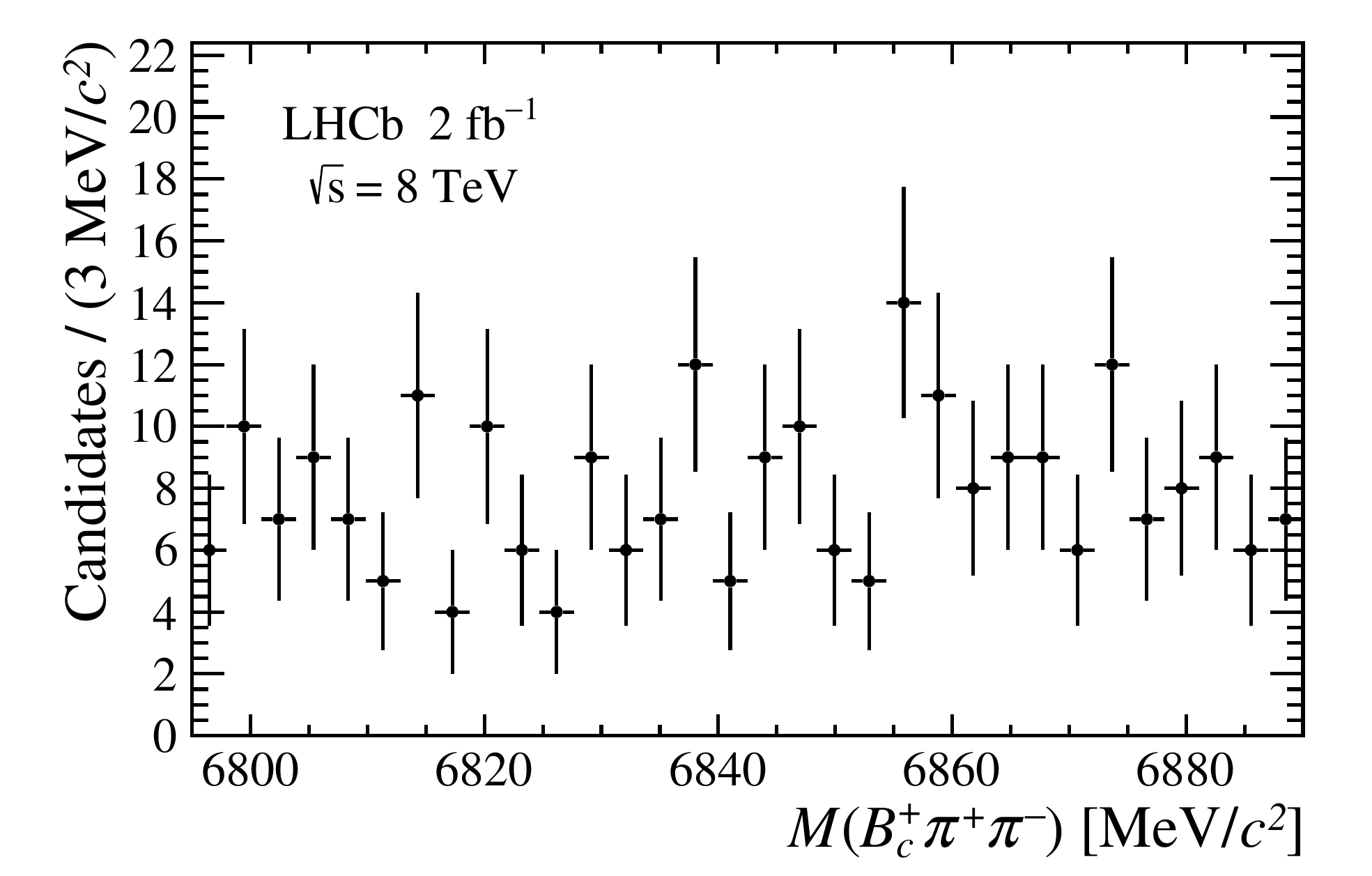} } \\
     \subfloat[MLP category: [0.4,0.6) ]{ \includegraphics[scale=0.39]{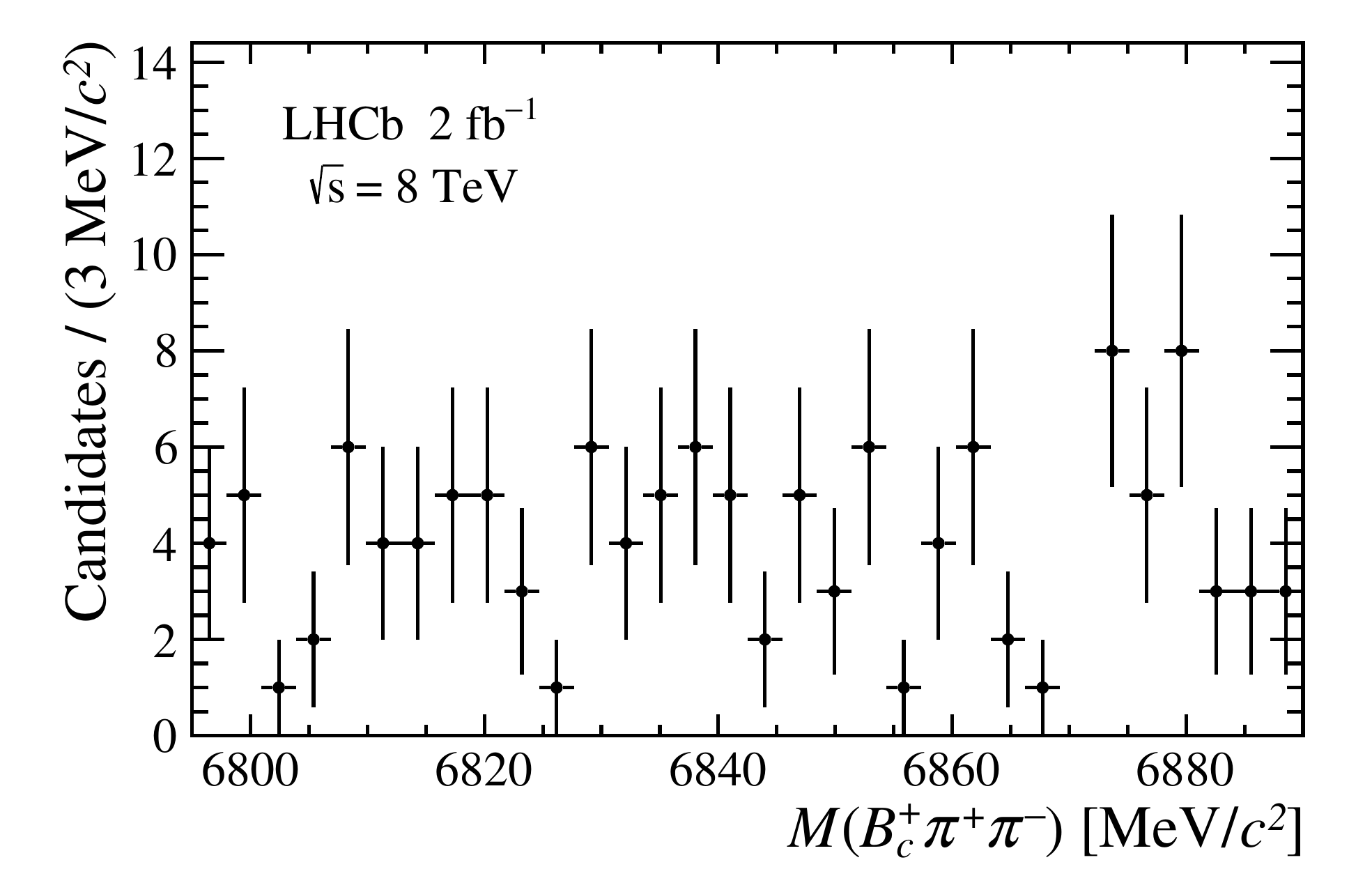} }
	  \subfloat[MLP category: {[0.6,1.0]} ]{ \includegraphics[scale=0.39]{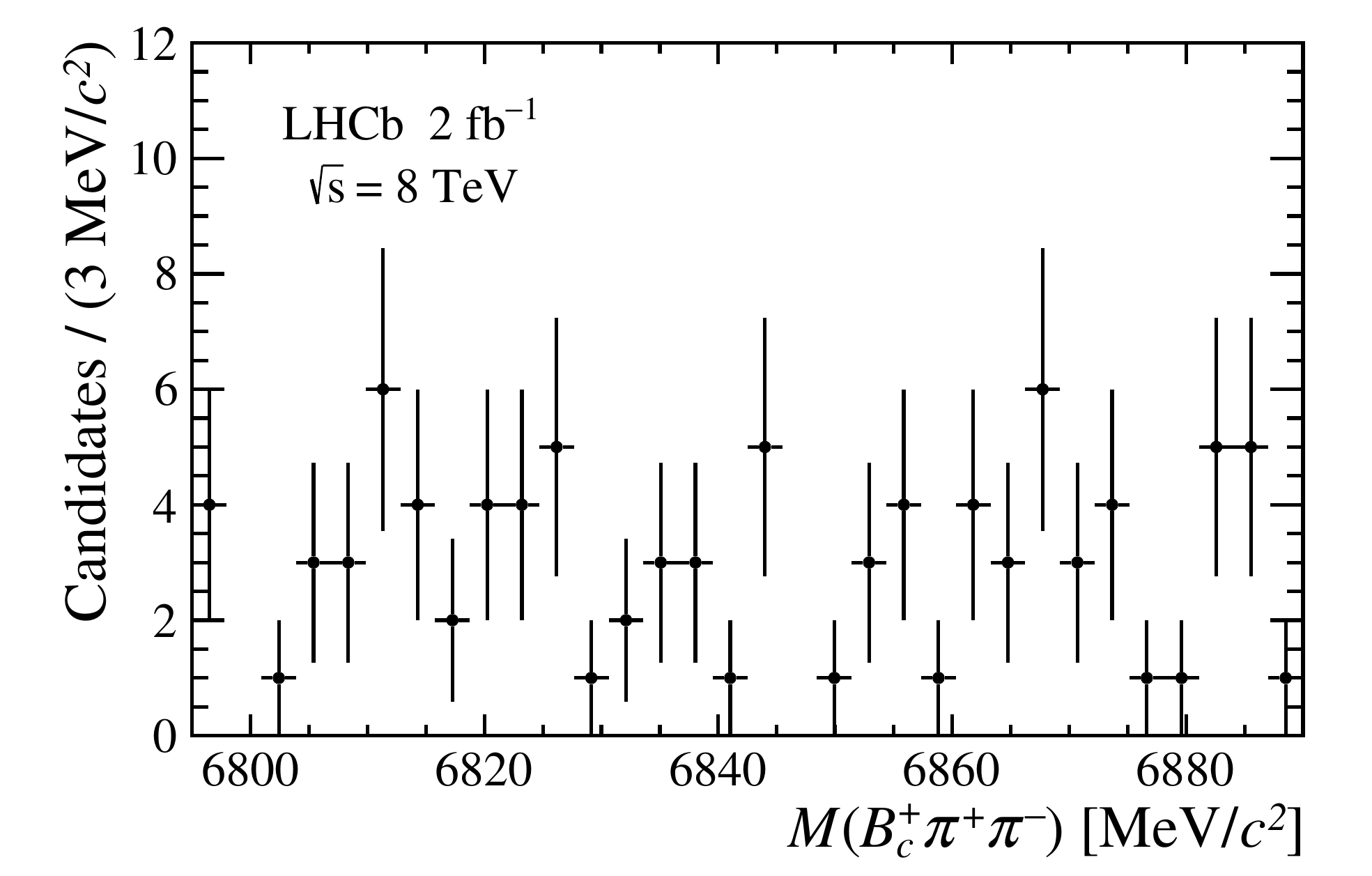} }
   \caption{\small
   Mass distributions of the selected $\Bcp\pip\pim$ candidates in the range $[6795,6890] \mevcc$ for the four MLP categories.
   }
   \label{fig:M_MLPcut}
 \end{center}
 \vskip-1.0cm
\end{figure}

To reconstruct the $\twoBctwos$ states,
the $\Bcp$ candidates with $\MJpsipi \in [6200,6340] \mevcc$ are combined with two opposite-charge tracks.
The tracks are required to have $\pt > 0.25\gevc$, momenta larger than $2\gevc$ and good track-fit quality, and to be identified as pions. 
The $\twoBctwos$ candidates are required to have good $\Bcp\pip\pim$ vertex-fit quality.
To improve the $\twoBctwos$ mass resolution,
the mass of $\Bcp$ candidates is constrained to the known $\Bcp$ mass~\cite{LHCb-PAPER-2012-028}, 
and the reconstructed $\twoBctwos$ mesons are constrained to originate from the associated PV.
To optimise the sensitivity of the analysis,
a selection based on a multilayer perceptron~(MLP)~\cite{Hocker:2007ht} classifier is applied. 
To distinguish the signal candidates from combinatorial background,
the MLP classifier uses information on
the angles between the $\Bcp$ and $\pip$, $\Bcp$ and $\pim$, and $\pip$ and $\pim$ candidate momenta projected in the plane transverse to the beam axis;
the angles between the $\twoBctwos$ momentum and the $\Bcp$, $\pip$, and $\pim$ momenta in the $\twoBctwos$ centre-of-mass frame;
the minimum cosine value of the angles 
between the momentum of the $\Bcp$ meson or of one of the pions from $\twoBctwos$ and the momentum of the muons or pion from the $\Bcp$ meson; 
and the vertex-fit $\chi^2$ of the $\twoBctwos$ meson.
In simulation, these variables have similar distributions for the $\Bctwosdecay$ and
$\Bctwosstardecay$ decays.
Therefore, 
the combination of the simulated candidates for the decays $\Bctwosdecay$ and
$\Bctwosstardecay$ is used as signal for the MLP training,
and the background sample consists of the candidates in the lower and upper
sidebands of the $\MBcpipi$ mass spectrum in data, 
with $\MBcpipi \in [6555,6785]\mevcc$ and $[6900,7500]\mevcc$, respectively. 
The MLP response is transformed to make the signal candidates distributed 
evenly between zero and unity, 
and the background candidates cluster near zero.
Only the candidates with transformed output values smaller than 0.02 are rejected, 
retaining 98\% of the signal.
The remaining candidates are divided into four categories with the MLP response falling in
$(0.02,0.2)$, $[0.2,0.4)$, $[0.4,0.6)$ and $[0.6,1.0]$, respectively.
The $\MBcpipi$ distributions in the expected signal region for the four MLP categories are shown
in Fig.~\ref{fig:M_MLPcut}.
The mass resolutions on $\MBcpipi$ for the $\twoBctwos$ state, $\sigma_{\rm w}(\twoBctwos)$,
can be determined from the simulated samples of the $\Bctwosdecay$ and $\Bctwosstardecay$ decays.
The differences between the mass resolutions in data and simulation 
are evaluated with the control decay mode $\Bcp \to \jpsi \pip\pim\pip$, 
which has the same final state as the signal and a large yield,
and are corrected by applying a scale factor.
The obtained mass resolutions are
$\sigma_{\rm w}(\Bctwos) = 2.05 \pm 0.05 \mevcc$
and
$\sigma_{\rm w}(\Bctwosstar) = 3.17 \pm 0.03 \mevcc$.
The $\MBcpipi$ distributions are consistent with the background-only hypothesis, 
as determined by the scan described below.

%% file: upperlim.tex
\section{Upper limits}
\label{sec:upperlim}
As no significant $\twoBctwos$ signal is found, 
upper limits are set, for each $\twoBctwos$ mass hypothesis,
on the ratio $\mathcal{R}$
of the $\twoBctwos$ production cross-section times the
branching fraction of $B_c^{(*)}(2S)^{+} \to \B_c^{(*)+} \pip\pim$ to 
the production cross-section of the $\Bcp$ state.
The ratio $\mathcal{R}$ is determined for $\twoBctwos$ and $\Bcp$ candidates in the kinematic ranges $\pt \in [0,20]\gevc$ and rapidity~$y \in [2.0,4.5]$, 
and is expressed as
\begin{eqnarray}
   \label{eq:R}
  \begin{aligned}
    \mathcal{R}  &= \frac{\sigma_{B_c^{(*)}(2S)^{+}}}{\sigma_{\Bcp}} \cdot
    \BR(B_c^{(*)}(2S)^{+} \to \B_c^{(*)+} \pip\pim) \\
    & = \frac{N_{B_c^{(*)}(2S)^{+}}}{N_{\Bcp}} \cdot
    \frac{\varepsilon_{\Bcp}}
    {\varepsilon_{B_c^{(*)}(2S)^{+}}},
  \end{aligned}
\end{eqnarray}
where
$\sigma$ is the production cross-section,
$N$ the yield,
and $\varepsilon$ the efficiency of reconstructing and selecting the $\Bcp$ or $\twoBctwos$ candidates
in the required $\pt$ and $y$ regions.
In the case $\Delta M = 0$, 
the reconstructed $\Bctwos$ and $\Bctwosstar$ states fully overlap,
and the ratio $\mathcal{R}$ corresponds to the sum of the $\mathcal{R}$ values of the $\Bctwos$ and $\Bctwosstar$ states.
The upper limits are calculated using the ${\rm CL_{s}}$ method~\cite{Read:2002hq},
in which the upper limit for each mass hypothesis is obtained from the ${\rm CL_{s}}$ value calculated as a function of the ratio $\mathcal{R}$.
The test statistic is the ratio of the likelihoods of the signal-plus-background hypothesis and the background-only hypothesis,
defined as
\begin{equation}
   \label{eq:calQ}
   \mathcal{Q}(N_{\rm obs};N_S,N_B) = \frac{\mathcal{L}(N_{\rm obs};N_S+N_B)}{\mathcal{L}(N_{\rm obs};N_B)},
\end{equation}
where $N_{\rm obs}$ is the number of observed candidates,
$N_B$ is the expected background yield,
and $N_S$ is the expected signal yield.
For a given value of the ratio $\mathcal{R}$, $N_S$ is determined as
\begin{eqnarray}
   \label{eq:Ns}
  \begin{aligned}
    N_S = \mathcal{R} \cdot N_{\Bcp} \cdot   
    \frac{\varepsilon_{B_c^{(*)}(2S)^{+}}}
    {\varepsilon_{\Bcp}}.
  \end{aligned}
\end{eqnarray}
The likelihood $\mathcal{L}$ is defined as  
\begin{equation}
   \mathcal{L}(n;x) = \frac{e^{-x}}{n!} x^{n}.
\end{equation}
The total statistical test value $\mathcal{Q}_{\rm tot}$ is the product of that for each of the four MLP categories.
The ${\rm CL_{s}}$ value is the ratio of ${\rm CL_{s+b}}$ to ${\rm CL_{b}}$,
where ${\rm CL_{s+b}}$ is the probability to find a $\mathcal{Q}_{\rm tot}$ value smaller than the $\mathcal{Q}_{\rm tot}$ value found in the data sample under the signal-plus-background hypothesis,
and ${\rm CL_{b}}$ is equivalent probability under the background-only hypothesis.
The ${\rm CL_{s+b}}$ and ${\rm CL_{b}}$ values are obtained from pseudoexperiments,
in which the input variables are varied within their statistical and systematic uncertainties.
The $\Bctwos$ state is searched for by scanning the mass region 
\mbox{$\MBcpipi \in [6830,6890] \mevcc$},
which is motivated by theoretical predictions~\cite{Gershtein:1987jj, Chen:1992fq, Eichten:1994gt, Kiselev:1994rc,
  Gupta:1995ps, Fulcher:1998ka, Ebert:2002pp, Godfrey:2004ya, Wei:2010zza, Rai:2006dt, El:2005}.
The value of $\Delta M$ is successively fixed to $0$, $15$, $25$ and $35 \mevcc$. 
The search windows are within $\pm 1.4\sigma_{\rm w}(\twoBctwos)$ of the $\twoBctwos$ mass hypotheses.
This choice of the search window gives the best sensitivity according to Ref.~\cite{Tisserand:1997av}.

\begin{figure}[hb]
  \vskip-0.5cm
  \begin{center}
     \subfloat[MLP category: (0.02,0.2) ]{ \includegraphics[scale=0.39]{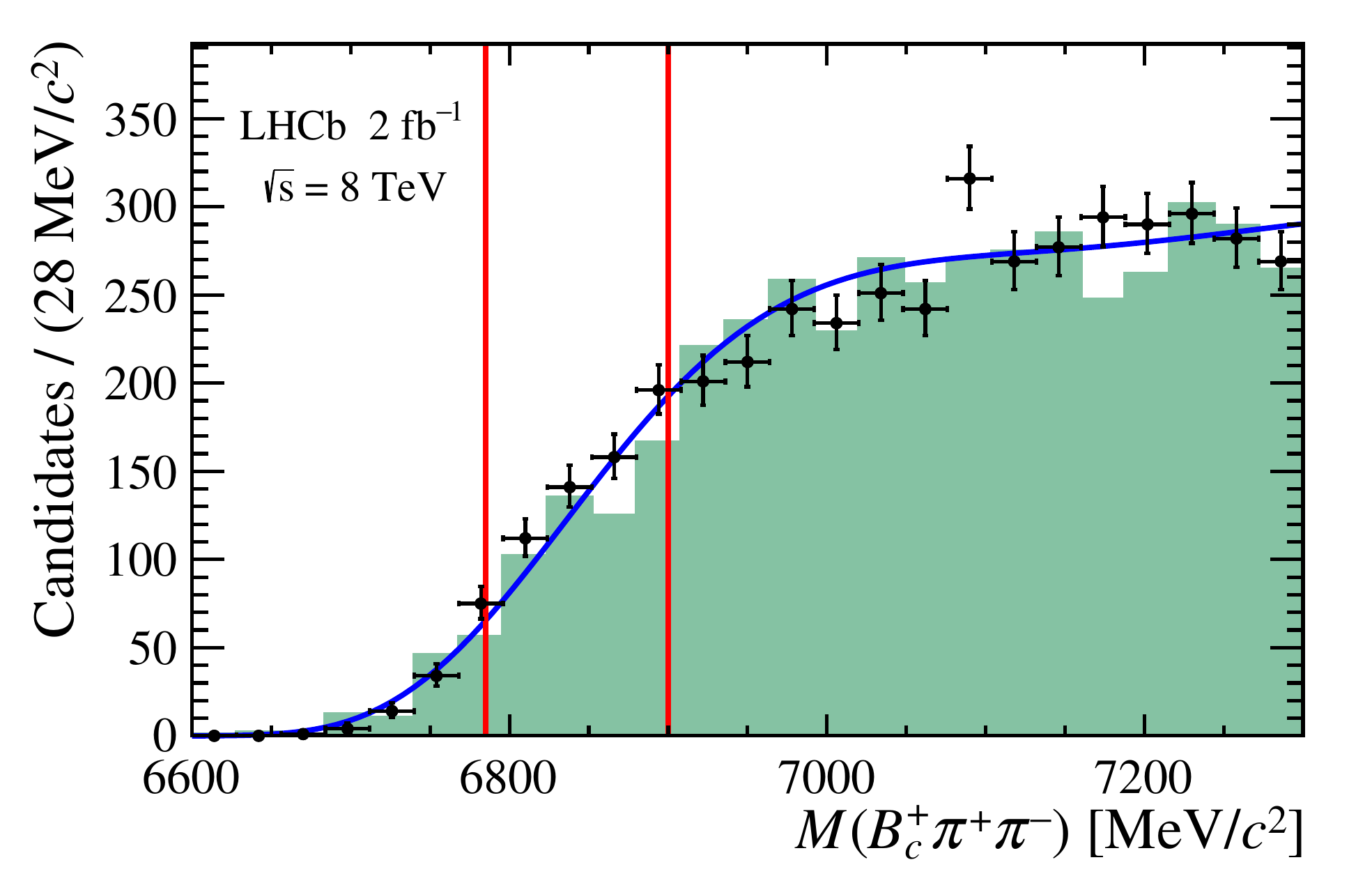} }
     \subfloat[MLP category: [0.2,0.4) ]{ \includegraphics[scale=0.39]{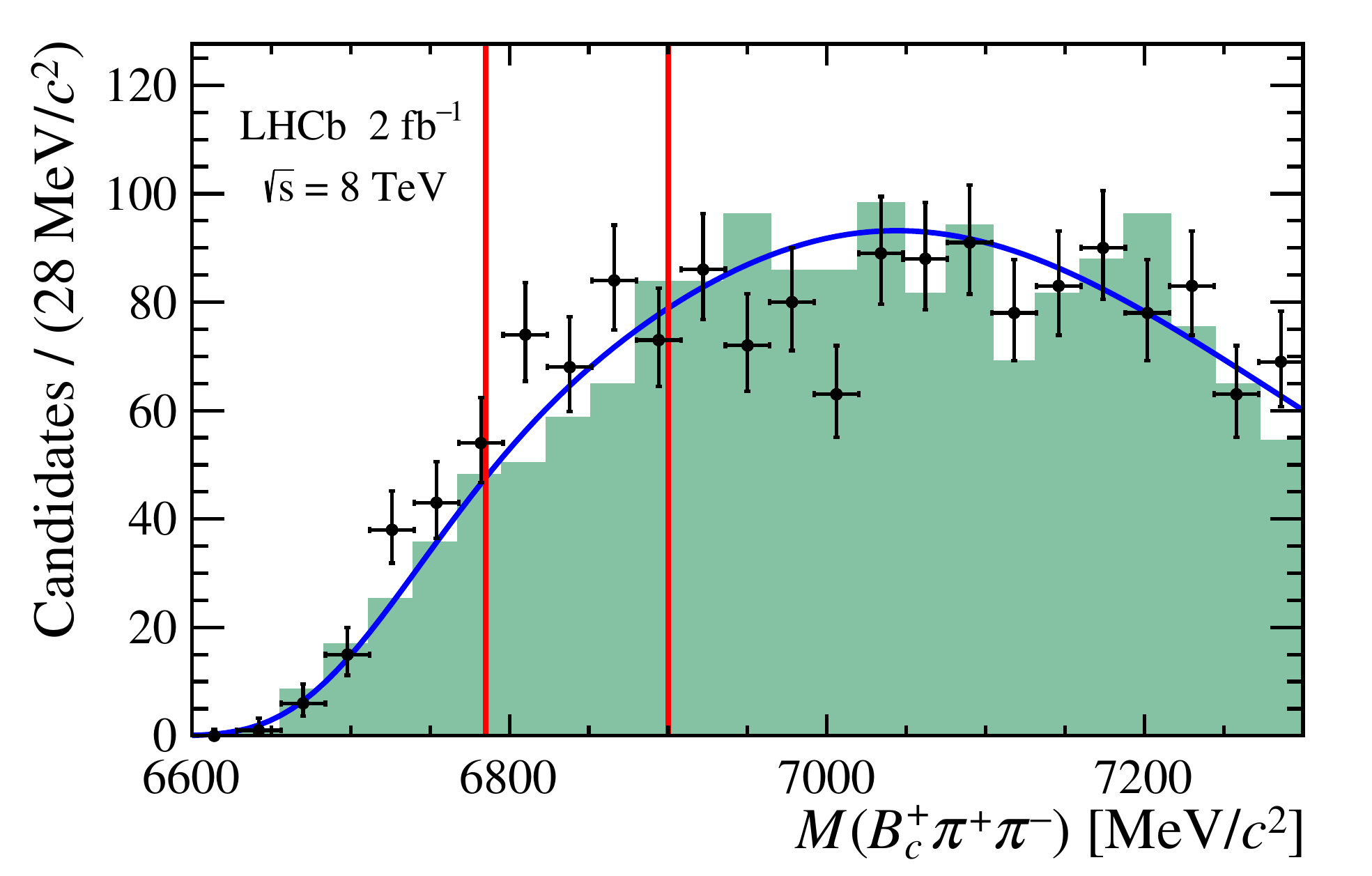} } \\
     \subfloat[MLP category: [0.4,0.6) ]{ \includegraphics[scale=0.39]{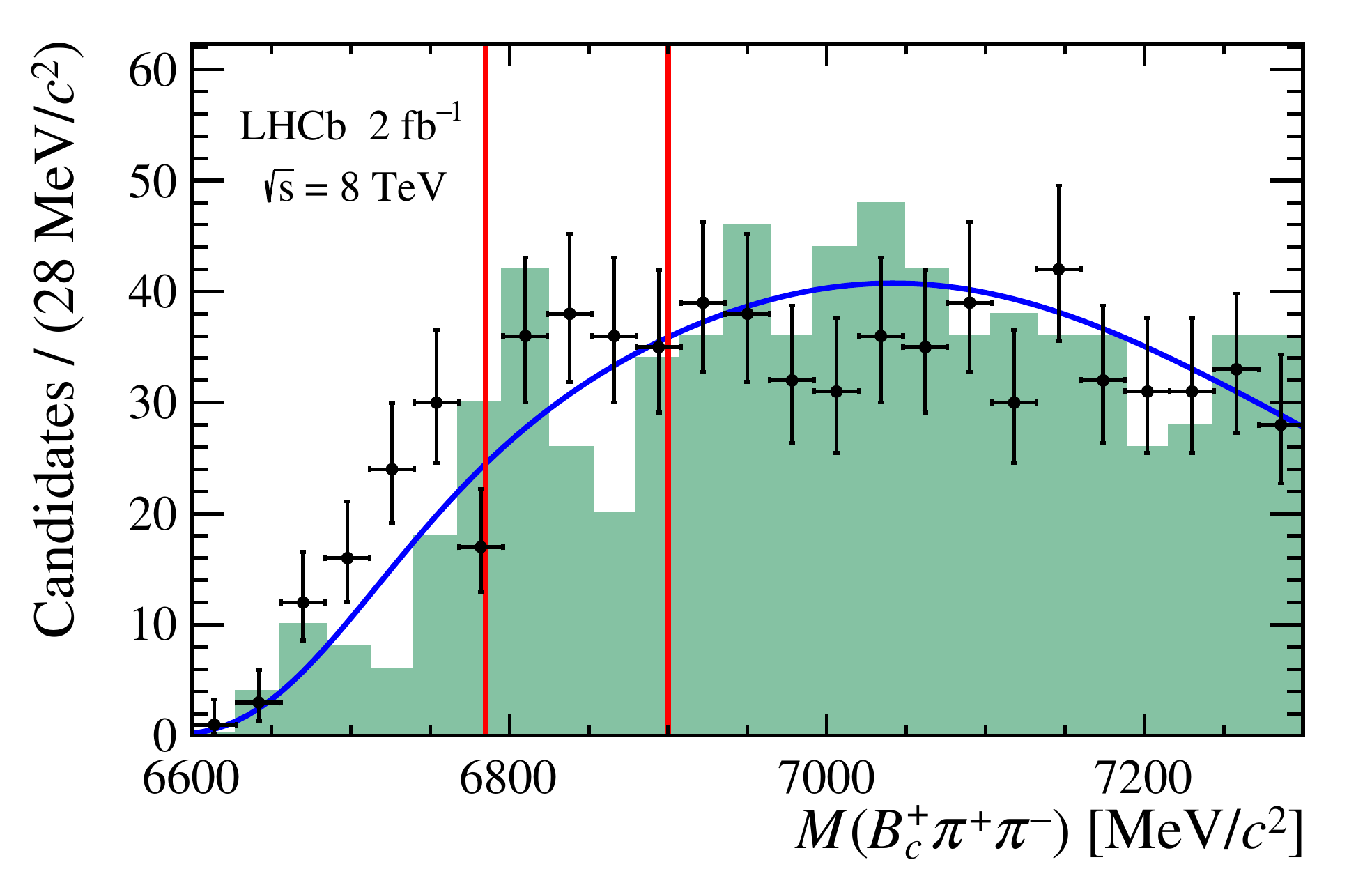} }
	  \subfloat[MLP category: {[0.6,1.0]} ]{ \includegraphics[scale=0.39]{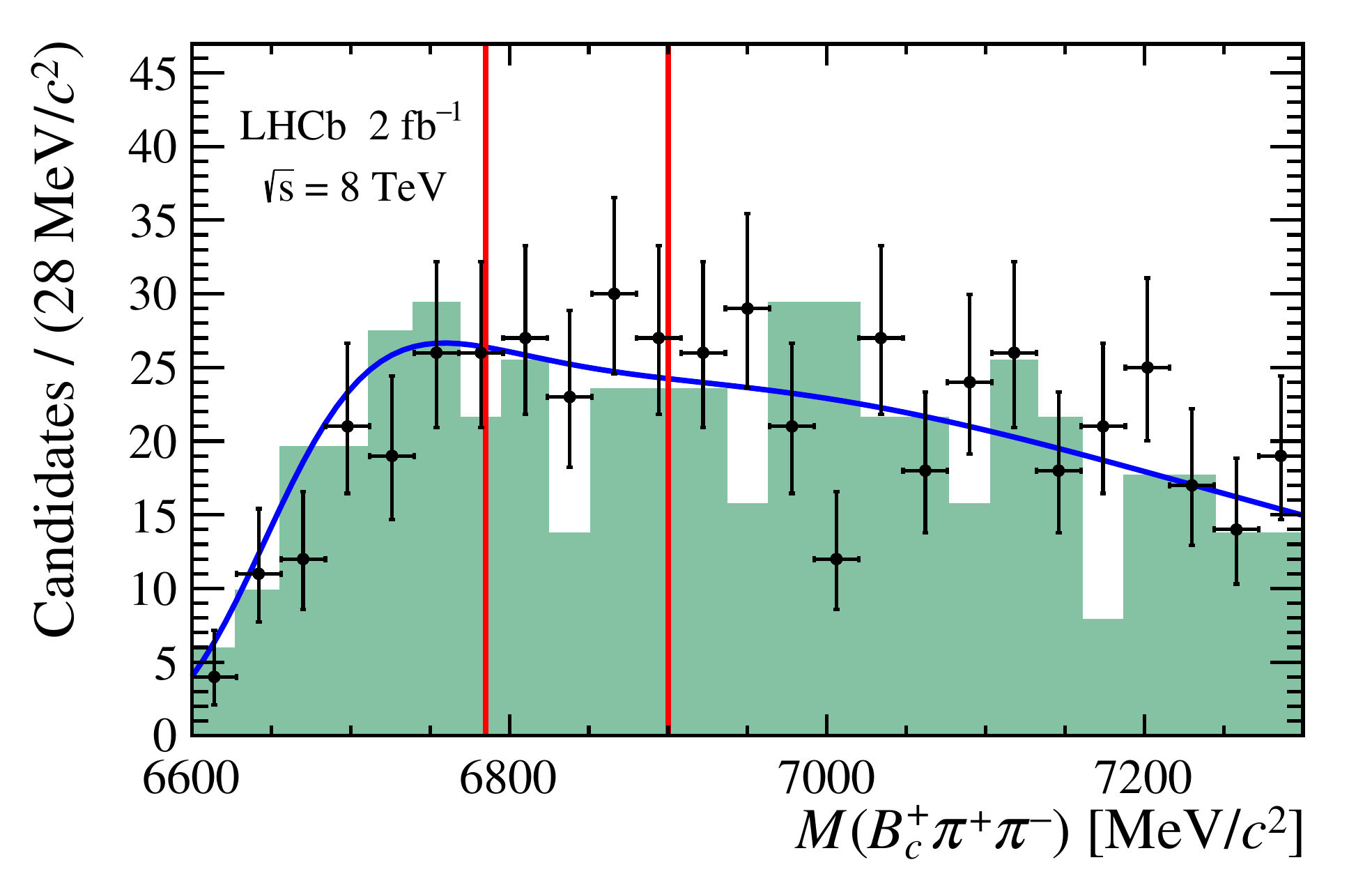} }
   \caption{The $\MBcpipi$ distributions in the same-sign~(darkgreen shaded areas) and data~(points with error bars) samples in the range $[6600,7300] \mevcc$ with the background model~(blue solid line) overlaid,
   for the four MLP categories.
   The areas between the two vertical red lines are the signal regions.}
   \label{fig:bkg}
 \end{center}
 \vskip-1.0cm
\end{figure}

The selection efficiencies $\varepsilon_{\Bcp}$ and $\varepsilon_{B_c^{(*)}(2S)^{+}}$
are estimated using simulation.
The track reconstruction efficiency is studied in a data control sample of $\jpsi \to \mup\mun$ decays using a tag-and-probe technique~\cite{LHCb-DP-2013-002},
in which one of the muons is fully reconstructed as the tag track, 
and the other muon, the probe track, is reconstructed using only information from the TT detector and the muon stations. 
The track reconstruction efficiency is the fraction of $\jpsi$ candidates whose probe tracks match fully reconstructed tracks.
The particle-identification~(PID) efficiency of the two opposite-charge pions is determined with a data-driven method,
using a $\pip$ sample from $D^{*}$-tagged $D^0 \to \Km \pip$ decays. 
The total efficiency $\varepsilon_{\Bcp}$ is determined to be $0.0931\pm0.0005$,
where the uncertainty is the statistical uncertainty of the simulated sample.
The $\twoBctwos$ efficiencies obtained from the default simulation,
where $\MBctwos = 6858\mevcc$ and $\MBctwosstar = 6890\mevcc$,
are summarised in Table~\ref{tab:eff}.
The variation of the efficiencies with respect to $\MBctwos$ and $\MBctwosstar$, 
assumed to be linear, 
is studied using the data simulated with different mass settings. 
This variation is considered when searching for the $\twoBctwos$ states at other masses.
The expected background yield in each of the $\twoBctwos$ signal regions, $N_B$, is estimated via extrapolation from the $\MBcpipi$ sidebands
for each MLP category.
The background is modelled by an empirical threshold function as shown in Fig.~\ref{fig:bkg},
where the threshold is taken to be $\MBcp+\Mpip+\Mpim=6555\mevcc$.
The other parameters are fixed according to the $\MBcpipi$ distribution of the same-sign sample,
which is constructed with $\Bcp \pip \pip$ or $\Bcp \pim \pim$ combinations.

\begin{table}
\begin{center}
\caption{\small \label{tab:eff}
Efficiencies for the $\twoBctwos$ states in the regions $\pt \in [0,20]\gevc$ and $y \in [2.0,4.5]$ for each MLP category. 
The efficiencies obtained before applying the MLP classifier are $0.0091 \pm 0.0002$ and $0.0086 \pm 0.0001$ for $\Bctwos$ and $\Bctwosstar$, respectively.
The uncertainties are  statistical only, and are due to the limited size of the simulated sample.}
\scalebox{0.9}{
\begin{tabular}{@{}ccccc@{}}
\toprule
MLP category & $(0.02,0.2)$ & $[0.2,0.4)$ & $[0.4,0.6)$ & $[0.6,1.0]$ \\
 & \multicolumn{4}{c}{Efficiencies in \%} \\
\midrule
$\Bctwos$ & $0.148 \pm 0.006$ & $0.140 \pm 0.006$  & $0.130 \pm 0.006$ & $0.256 \pm 0.008$ \\ 
$\Bctwosstar$ & $0.118 \pm 0.003$ & $0.140 \pm 0.004$  & $0.144 \pm 0.004$ & $0.288 \pm 0.005$ \\ 
\bottomrule
\end{tabular}
}
\end{center}
\end{table}

The sources of systematic uncertainties that affect the upper limit calculation are studied and summarised in Table~\ref{tab:sys}.
The systematic uncertainty on $N_{\Bcp}$ comes from the potentially imperfect
modelling of the signal,
and has been studied using pseudoexperiments.
The uncertainty on $\varepsilon_{\Bcp}$ is due to the limited size of the simulated sample.
The uncertainty on $N_B$ comes both from differences between the
combinatorial backgrounds in the opposite-sign and the same-sign data samples
and from the potential mismodelling of the background. 
The former is studied by performing a large set of pseudoexperiments,
in which the samples are generated by randomly taking candidates from the data sample, 
while the candidates in $\MBcpipi \in [6785,6900] \mevcc$ are taken from the same-sign sample. 
The $\MBcpipi$ distributions of the pseudosamples are fit using the same function as in the nominal background modelling.
The difference between the mean value of $N_B$ obtained from the pseudoexperiments 
and the nominal value is taken as the systematic uncertainty.
The potential mismodelling of the background is estimated by using 
the Bukin function~\cite{Silagadze2007} as an alternative model
and the differences to the nominal results are taken as systematic uncertainties.
The uncertainties on $\varepsilon_{\twoBctwos}$ are dominated by the uncertainty 
due to the finite size of the simulated samples,
but also include the systematic uncertainties on the PID and track reconstruction efficiency calibration,
which come from the limited size and the binning scheme of the calibration samples.
The variations of efficiency with respect to $\MBctwos$ and $\MBctwosstar$ are fitted with linear functions, 
and the uncertainties of such fits are taken as systematic uncertainties.

\begin{table}
\begin{center}
\caption{\small \label{tab:sys}
Summary of the systematic uncertainties entering the upper limit calculation for the four MLP categories.}
\scalebox{0.9}{
\begin{tabular}{@{}lcccc@{}}
\toprule
MLP category & $(0.02,0.2)$ & $[0.2,0.4)$ & $[0.4,0.6)$ & $[0.6,1.0]$ \\
\midrule
$N_{\Bcp}$ & \multicolumn{4}{c}{1.0\%} \\
$\varepsilon_{\Bcp}$ & \multicolumn{4}{c}{0.5\%} \\
$N_B$ & $4.2\%$ & $9.0\%$ & $15.0\%$ & $6.9\%$ \\
\midrule
\multicolumn{5}{l}{$\Bctwosdecay$} \\
\midrule
$\varepsilon_{\Bctwos}$ & $4.6\%$ & $4.7\%$ & $4.9\%$ & $3.6\%$ \\
Efficiency variation \vs $\MBctwos$ & $0.6\%$ & $1.3\%$ & $1.8\%$ & $2.7\%$ \\
\midrule
\multicolumn{5}{l}{$\decay{\Bctwosstar}{\Bcstar\pip\pim}$} \\
\midrule
$\varepsilon_{\Bctwosstar}$ & $3.5\%$ & $3.3\%$ & $3.3\%$ & $2.7\%$ \\
Efficiency variation \vs $\MBctwosstar$ & $1.0\%$ & $1.8\%$ & $2.5\%$ & $4.3\%$ \\
\bottomrule
\end{tabular}
}
\end{center}
\end{table}

\begin{figure}
 \begin{center}
    \subfloat[$\Delta M = 0 \mevcc$]{ \includegraphics[scale=0.39]{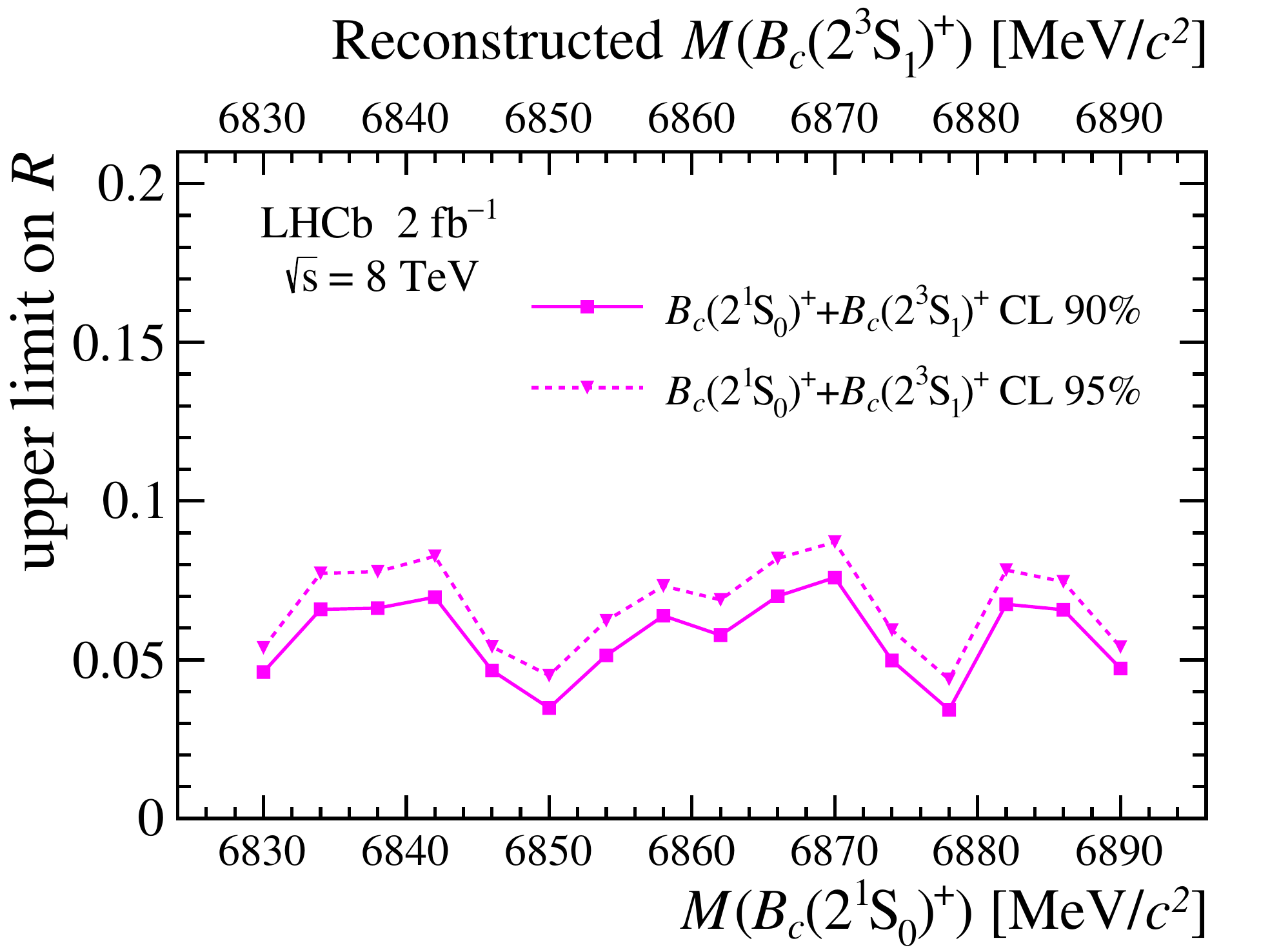} }
    \subfloat[$\Delta M = 15 \mevcc$]{ \includegraphics[scale=0.39]{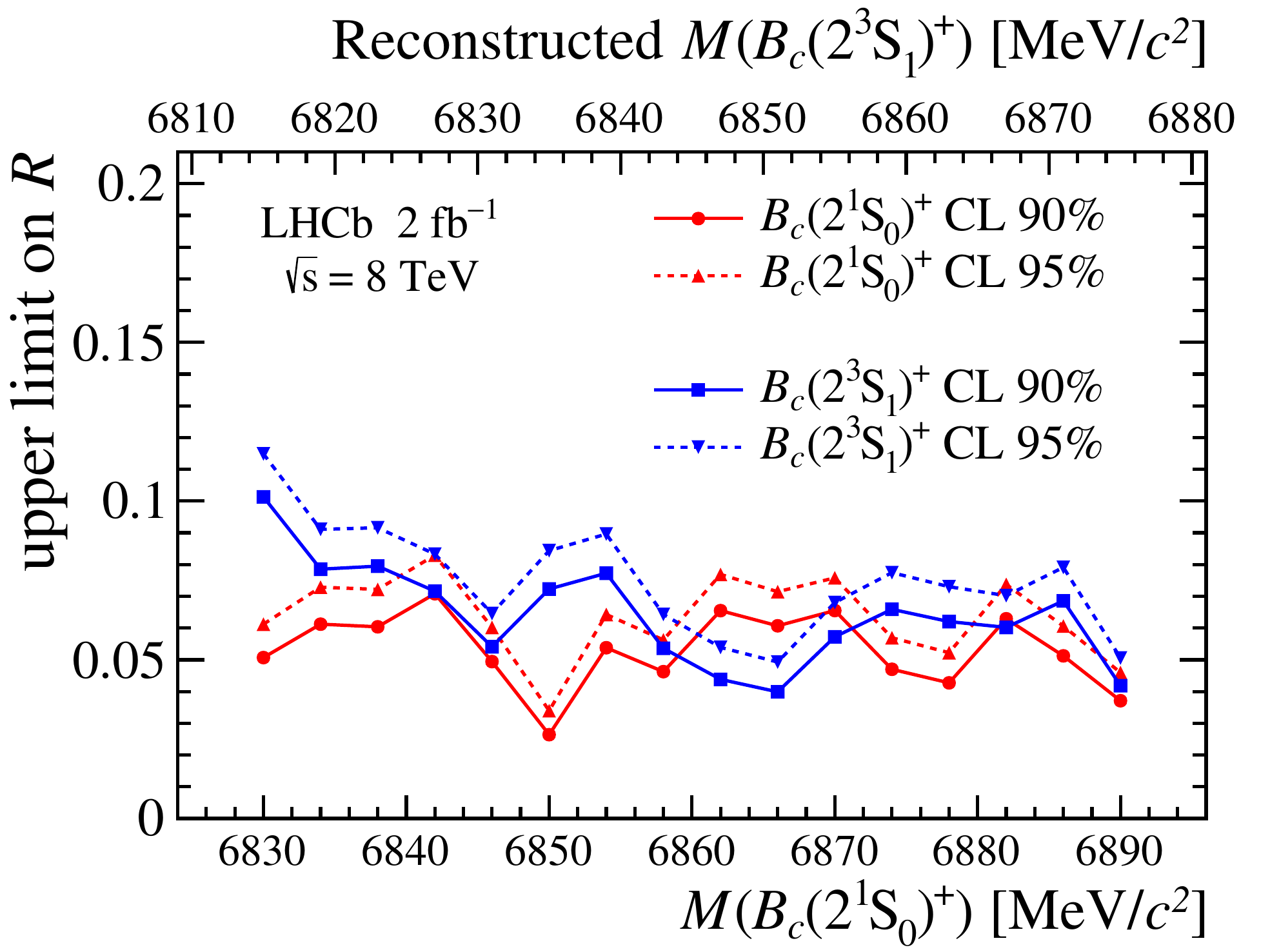} } \\
    \subfloat[$\Delta M = 25 \mevcc$]{ \includegraphics[scale=0.39]{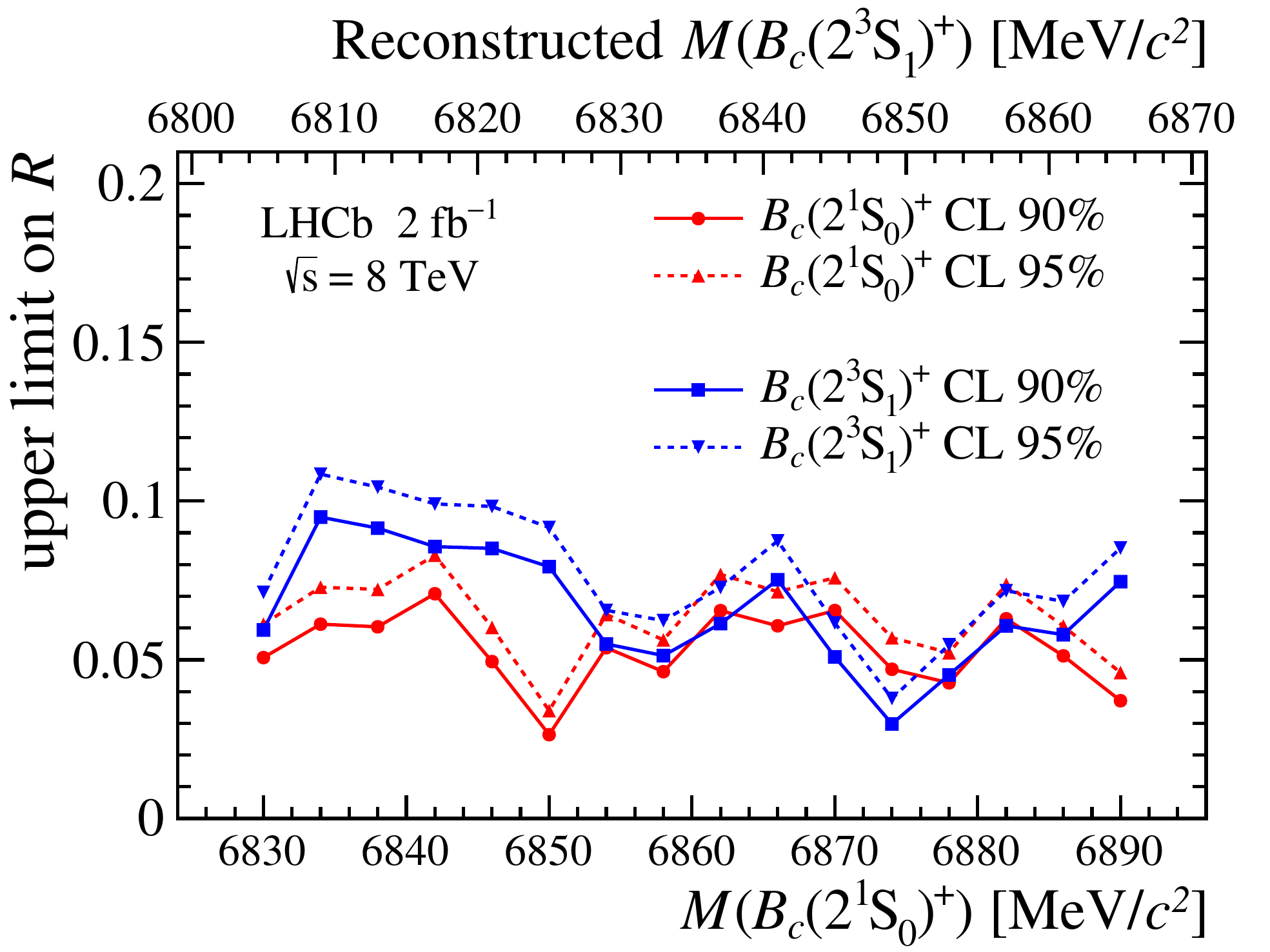} }
    \subfloat[$\Delta M = 35 \mevcc$]{ \includegraphics[scale=0.39]{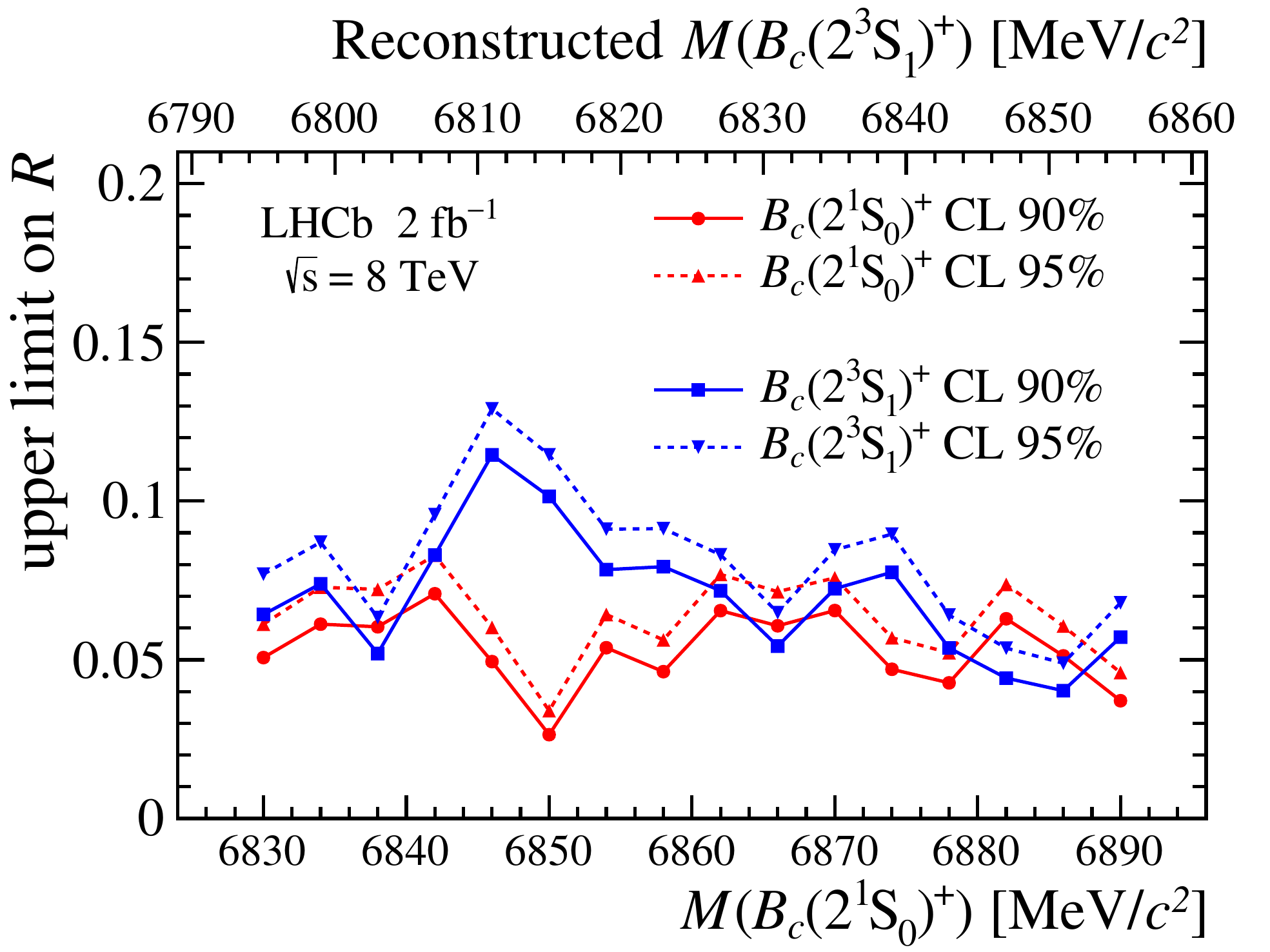} }
    \caption{The upper limits on the ratio $\mathcal{R}(\twoBctwos)$ at $95\%$ and $90\%$ confidence levels under different mass splitting $\Delta M$ hypotheses.
    }
   \label{fig:uplimsum}
 \end{center}
\end{figure}

No evidence of the $\twoBctwos$ signal is observed. 
The measurement is consistent with the background-only hypothesis
for all mass assumptions.
The upper limits at $90\%$ and $95\%$ confidence levels~(CL)
on the ratio $\mathcal{R}$, as functions of the $\twoBctwos$ mass states, 
are shown in Fig.~\ref{fig:uplimsum}.
All the upper limits at 95\% CL on the ratio $\mathcal{R}$ are contained between $0.02$ and $0.14$.
Theoretical models predict that the ratio $\mathcal{R}$ has no significant dependence on $y$ and $\pt$ of the $\Bcp$ mesons~\cite{Chang:2005hq},
allowing comparison with the \atlas result~\cite{Aad:2014laa}.
The most probable interpretation of the \atlas measurement is that
it is either the $\Bctwosstar$ state or a sum of $\Bctwos$ and $\Bctwosstar$ signals under the $\Delta M \sim 0$ scenario.
For both interpretations of the \atlas measurement, 
the comparison of the ratio $\mathcal{R}$ between the \lhcb upper limits in the vicinity of the peak claimed by \atlas at $M(B_c^{(*)}(2S)^+)=6842~\mevcc$
and the ratios determined by \atlas are given in Table~\ref{tab:result}. 
The \lhcb and \atlas results are compatible only in case of very large~(unpublished) relative efficiency of reconstructing the $\twoBctwos$ candidates with respect to the $\Bcp$ signals for the \atlas measurement.

\begin{table}
\begin{center}
\caption{\small \label{tab:result}
Comparison of the $\mathcal{R}$ value between the \lhcb upper limits at 95\% CL and the \atlas measurement~\cite{Aad:2014laa},
where $0 < \varepsilon_{7,8} \leq 1$ are the relative efficiencies of reconstructing the $\twoBctwos$ candidates with respect to the $\Bcp$ signals
for the 7 and $8 {\ensuremath{\mathrm{\,Te\kern -0.1em V}}\xspace}$ data, respectively.
}
\scalebox{0.9}{
\begin{tabular}{@{}ccc@{}}
\toprule
 & $\sqs = 7\tev$ & $\sqs = 8\tev$ \\
 \midrule
\atlas & $(0.22 \pm 0.08\stat) / \varepsilon_{7}$ & $(0.15 \pm 0.06\stat) / \varepsilon_{8}$ \\
\midrule
\lhcb & -- & $<[0.04,0.09]$\\
\bottomrule
\end{tabular}
}
\end{center}
\end{table}

\section{Summary}
\label{sec:summary}
In summary, a search for the \Bctwos and \Bctwosstar states 
is performed at LHCb 
with a data sample of $pp$ collisions, 
corresponding to an integrated luminosity of $2\invfb$,
recorded at a centre-of-mass energy of 8 TeV. 
No significant signal is found.
Upper limits on the \Bctwos and \Bctwosstar production cross-sections times the
branching fraction of $\twoBctwosdecay$
relative to the $\Bc$ cross-section, 
are given as a function of the $\Bctwos$ and $\Bctwosstar$ masses.

%% file: acknowledgements.tex
\section*{Acknowledgements}

\noindent 
We thank Chao-Hsi~Chang and Xing-Gang~Wu for frequent and interesting discussions on the production of the $B_c$ mesons.
We express our gratitude to our colleagues in the CERN
accelerator departments for the excellent performance of the LHC. We
thank the technical and administrative staff at the LHCb
institutes. We acknowledge support from CERN and from the national
agencies: CAPES, CNPq, FAPERJ and FINEP (Brazil); MOST and NSFC
(China); CNRS/IN2P3 (France); BMBF, DFG and MPG (Germany); INFN
(Italy); NWO (The Netherlands); MNiSW and NCN (Poland); MEN/IFA
(Romania); MinES and FASO (Russia); MinECo (Spain); SNSF and SER
(Switzerland); NASU (Ukraine); STFC (United Kingdom); NSF (USA).  We
acknowledge the computing resources that are provided by CERN, IN2P3
(France), KIT and DESY (Germany), INFN (Italy), SURF (The
Netherlands), PIC (Spain), GridPP (United Kingdom), RRCKI and Yandex
LLC (Russia), CSCS (Switzerland), IFIN-HH (Romania), CBPF (Brazil),
PL-GRID (Poland) and OSC (USA). We are indebted to the communities
behind the multiple open-source software packages on which we depend.
Individual groups or members have received support from AvH Foundation
(Germany), EPLANET, Marie Sk\l{}odowska-Curie Actions and ERC
(European Union), ANR, Labex P2IO, ENIGMASS and OCEVU, and R\'{e}gion
Auvergne-Rh\^{o}ne-Alpes (France), RFBR and Yandex LLC (Russia), GVA,
XuntaGal and GENCAT (Spain), Herchel Smith Fund, the Royal Society,
the English-Speaking Union and the Leverhulme Trust (United Kingdom).

%% file: LHCb_Authorship_flat_09-Oct-2017.tex
\centerline{\large\bf LHCb collaboration}
\begin{flushleft}
\small
R.~Aaij$^{40}$,
B.~Adeva$^{39}$,
M.~Adinolfi$^{48}$,
Z.~Ajaltouni$^{5}$,
S.~Akar$^{59}$,
J.~Albrecht$^{10}$,
F.~Alessio$^{40}$,
M.~Alexander$^{53}$,
A.~Alfonso~Albero$^{38}$,
S.~Ali$^{43}$
% ...

{\footnotesize \it
% ...
$ ^{5}$Clermont Universit{\'e}, Universit{\'e} Blaise Pascal, CNRS/IN2P3, LPC, Clermont-Ferrand, France\\
% ...
$ ^{10}$Fakult{\"a}t Physik, Technische Universit{\"a}t Dortmund, Dortmund, Germany\\
% ...
$ ^{38}$ICCUB, Universitat de Barcelona, Barcelona, Spain\\
$ ^{39}$Instituto Galego de F{\'\i}sica de Altas Enerx{\'\i}as (IGFAE), Universidade de Santiago de Compostela, Santiago de Compostela, Spain\\
$ ^{40}$European Organization for Nuclear Research (CERN), Geneva, Switzerland\\
% ...
$ ^{43}$Nikhef National Institute for Subatomic Physics, Amsterdam, The Netherlands\\
% ...
$ ^{48}$H.H. Wills Physics Laboratory, University of Bristol, Bristol, United Kingdom\\
% ...
$ ^{53}$School of Physics and Astronomy, University of Glasgow, Glasgow, United Kingdom\\
% ...
$ ^{59}$University of Cincinnati, Cincinnati, OH, United States\\
% ...
}
\end{flushleft}